\documentclass[twocolumn]{elsarticle}
\usepackage{lipsum}
\makeatletter
\def\ps@pprintTitle{%
 \let\@oddhead\@empty
 \let\@evenhead\@empty
 \def\@oddfoot{}%
 \let\@evenfoot\@oddfoot}
\makeatother

\usepackage{lineno,hyperref}
\usepackage{units}                    
\usepackage{aas_macros}
\usepackage{amsmath}
\usepackage{amssymb}
\usepackage{tabularx}
\newcommand{\teff}{\ensuremath{\tau_{\rm eff}}}
\newcommand{\wdm}{{\rm wdm}} 
\graphicspath{{fig/}} 

\journal{Physics Letters B}

\begin{document}
\twocolumn[{
\begin{frontmatter}

\title{Cutoff in the Lyman-$\alpha$ forest power spectrum: warm IGM or
  warm dark matter?}
\author[{add1,add2}]{Antonella~Garzilli}
\cortext[mycorrespondingauthor]{Corresponding author}
\ead{garzilli@lorentz.leidenuniv.nl}
\author[add2]{Alexey~Boyarsky}
\author[add3]{Oleg~Ruchayskiy}
\address[add1]{GRAPPA, Institute of Theoretical Physics, University of Amsterdam, Science Park 904, 1090 GL Amsterdam}
\address[add2]{Lorentz Institute, Leiden University, Niels Bohrweg 2,
  Leiden, NL-2333 CA, The Netherlands}
\address[add3]{Discovery Center, Niels Bohr Institute, Blegdamsvej 17, DK-2100 Copenhagen, Denmark}
\begin{abstract}
We re-analyse high redshift and high resolution Lyman-$\alpha$ forest
spectra considered in \citep{Viel:2013apy}, seeking to constrain the
properties of warm dark matter particles.  Compared to this previous
work, we consider a wider range of thermal histories of the
intergalactic medium. We find that both warm and cold dark matter
models can explain the cut-off observed in the flux power spectra of
high-resolution observations equally well.  This implies, however,
very different thermal histories and underlying re-ionisation models.
We discuss how to remove this degeneracy.
\end{abstract}


\end{frontmatter}
}]

\section*{Introduction}  Dark matter is a central ingredient of the
current standard cosmological model. It drives the formation of
structures, and explains the masses of galaxies and galaxy clusters.
If dark matter is made of particles, these yet-unseen particles should
have been created in the early Universe long before the recombination
epoch. If such particles were relativistic at early times, they would
stream out from overdense regions, smoothing out primordial density
fluctuations. The signature of such \emph{warm dark matter} (WDM)
scenario would be the suppression of the matter power spectrum at
scales below their free-streaming horizon. From cosmological data at
large scales (CMB and galaxy surveys) we know that such a suppression
should be sought at comoving scales well below a Mpc.
 
The Lyman-$\alpha$ forest has been used for measuring the matter power
spectrum at such
scales~\cite{Croft:1997jf,McDonald:1999dt,Croft:2000hs}.  In previous
works only upper bounds had been reported on the mass of the thermal
relic
~\cite{Hansen:2001zv,Boyarsky:2008xj,Seljak:2006qw,Viel:2007mv,Viel:2006kd,Viel:2005qj}.
However, while in the SDSS spectra there is no cut-off in the
transmitted flux power spectrum, there is a cut-off in the high
resolution spectra, for example \cite{Croft:2000hs,Kim:2003qt,Seljak:2006qw}. Recently
\citep{Viel:2013apy} has observed the cut-off of the flux power
spectrum at scales $k \sim \unit[0.03]{s/km}$ and redshifts $z =
4.2-5.4$.

However, the Lyman-$\alpha$ forest method measures not the
distribution of dark matter itself, but only the neutral hydrogen
density as a proxy for the overall matter density.  The process of
reionization heats the hydrogen and prevents it from clustering at
small scales at the redshifts in question
\citep{gnedinhui98}. Therefore, the observed hydrogen distribution
eventually stops to follow the DM distribution. Indeed, it was
demonstrated in \citep{Viel:2013apy} that within $\Lambda$CDM
cosmology there exists a suitable thermal history of intergalactic
medium (IGM) that is consistent with the observed cutoff. This does
not mean, however, that this scenario is realised in nature.

In this Letter we investigate this issue in depth. We ask whether
\emph{the cutoff in the flux power spectrum can be attributed to the
  supression of small scales with warm dark matter} and what this
means for the thermal history of IGM.  To this end we reanalyze the
data used in~\cite{Viel:2013apy}. We use \emph{the same} suite of
hydrodynamical simulations of the IGM evolution with cold and warm
dark matter models as in~\cite{Viel:2013apy} and demonstrate that the
data is described equally well by the model, where flux power spectrum
suppression is mainly due to WDM.

\bigskip

\section*{Data and model}  The data set is constituted by 25
high-resolution quasar spectra, in the redshift interval $4.48\leq
z_{{\rm QSO}} \leq 6.42$. The spectra were taken with the Keck High
Resolution Echelle Spectrometer (HIRES) and the Magellan Inamory
Kyocera Echelle (MIKE) spectrograph on the Magellan clay telescope.
The QSO spectra are divided into four redshift bins centered on :
$z={}$4.2, 4.6, 5.0, 5.4. The resulting range of wave-numbers probed
by this dataset is $k = \unit[0.005-0.08]{s/km}$.
\bigskip

At these redshifts, the IGM is thought to be in a highly ionized
state, being photo-ionized and photo-heated by early sources.  Both
the WDM cosmology and the IGM temperature affect the amount of flux
power spectrum at small scales through three distinct physical
mechanisms: (1) a suppression in the initial matter power spectrum; (2) 
Jeans broadening; and (3) Doppler broadening of the absorption lines
\citep{gnedinhui98,theuns2000,Desjacques:2004xy,Peeples:2009uj,Garzilli:2015bha,
  kulkarni2015}. The first mechanism is cosmological, the latter two
are astrophysical. The Doppler broadening is a one dimensional
smoothing effect that originates from observing the hot IGM along a
line of sight. The Maxwellian distribution of velocities in
the gas then leads to the broadening effect. The Jeans broadening smoothes the
three-dimensional underlying 
gas distribution relative to the dark matter.

The level of ionization is captured by \emph{the effective optical
  depth}, $\tau_{\rm eff}$, that is computed from the mean flux,
$\langle F \rangle$, through the relation $\langle F(z)\rangle =
\exp(-\teff(z))$. Because the IGM spans a wide range of density,
describing the IGM temperature may be complicated in principle. But,
assuming that the IGM is heated by photo-heating, the temperature of
the IGM follows a simple power-law \emph{temperature-density
  relation}~\cite{Hui:1997dp}:
\begin{equation}
  \label{eq:1}
  T(\delta) = T_0(z) \bigl(1 + \delta\bigr)^{\gamma(z)-1}.
\end{equation}
where $\delta = \delta \rho_\mathrm{m}/\bar \rho_\mathrm{m}$ is the matter
overdensity and $T_0(z)$, $\gamma(z)$ are unknown functions of redshift.
The results of Ref.~\cite{Viel:2013apy} are based  on  single power-law
parametrizations, $T_0(z)$ and $\gamma(z) $.
In this letter we let the parameters of the IGM thermal
state vary independently in each redshift bin, with a total of 8
parameters describing the IGM thermal state ($T_0(z_i)$ and
$\gamma(z_i)$ in 4 distinct redshift intervals).\footnote{Ref.~\cite{Viel:2013apy} also performed such a ``binned
analysis'', see the detailed comparison below.}

We want to point out that $T_0$ and $\gamma$ are not varied in
post-processing. The original work of \citep{Viel:2013apy} considered
9 simulation runs with distinct thermal histories for each cosmology
considered. The different thermal histories are realized by changing
the photo-heating function in the simulations. The resulting values of
$T_0$ and $\gamma$ are approximately distributed on a regular grid. In
\citep{Viel:2013apy} the effect of Jeans smoothing is accounted by
considering two additional simulation runs, where the time at which
the ultraviolet background is switched on, $z_{\rm reion}$, is varied.
We caution the reader that the resulting constraints on $z_{\rm
  reion}$ must not be intended as a measurement of the time of
reionization, because this depends on the details of the
implementation of the ultraviolet background. Instead, varying $z_{\rm
  reion}$ must be considered as a way to account for the unknown level
of Jeans smoothing. Finally, as in \cite{Viel:2013apy}, we allow the
effective optical depth vary independently in each redshift bin,
$\tau_{\rm eff}[z_i]$.

It should be noted that this interpolation scheme between simulations
with different temperatures may also vary the amount of Jeans
broadening (also known as the ``filtering scale'').  While the
degeneracy between the WDM cosmologies and the Doppler smoothing has
been extensively considered in the literature, the degeneracy between
Jeans smoothing and WDM cosmology has not been considered in depth so
far. In particular this has not been done for the suite of simulations
in the original work \citep{Viel:2013apy} on which we base our
analysis.
We leave the study of the degeneracy between the Jeans smoothing and WDM for
future work.

The results also depend on the cosmological parameters $n_s, \,
\Omega_M,\, \sigma_8,\, H_0$. However the small scale Lyman-$\alpha$
data by itself does not sufficiently constrain the cosmological
parameters. Therefore, in the final likelihood function for these
parameters we used best fit Planck values \citep{planck2013} with
Gaussian priors (as in~\cite{Viel:2013apy}), $\Omega_M = 0.315 \pm
0.017$, $\sigma_8 = 0.829 \pm 0.013$, $n_s = 0.9603 \pm 0.0073$.

\begin{figure}
  \centering \includegraphics[width=\columnwidth]{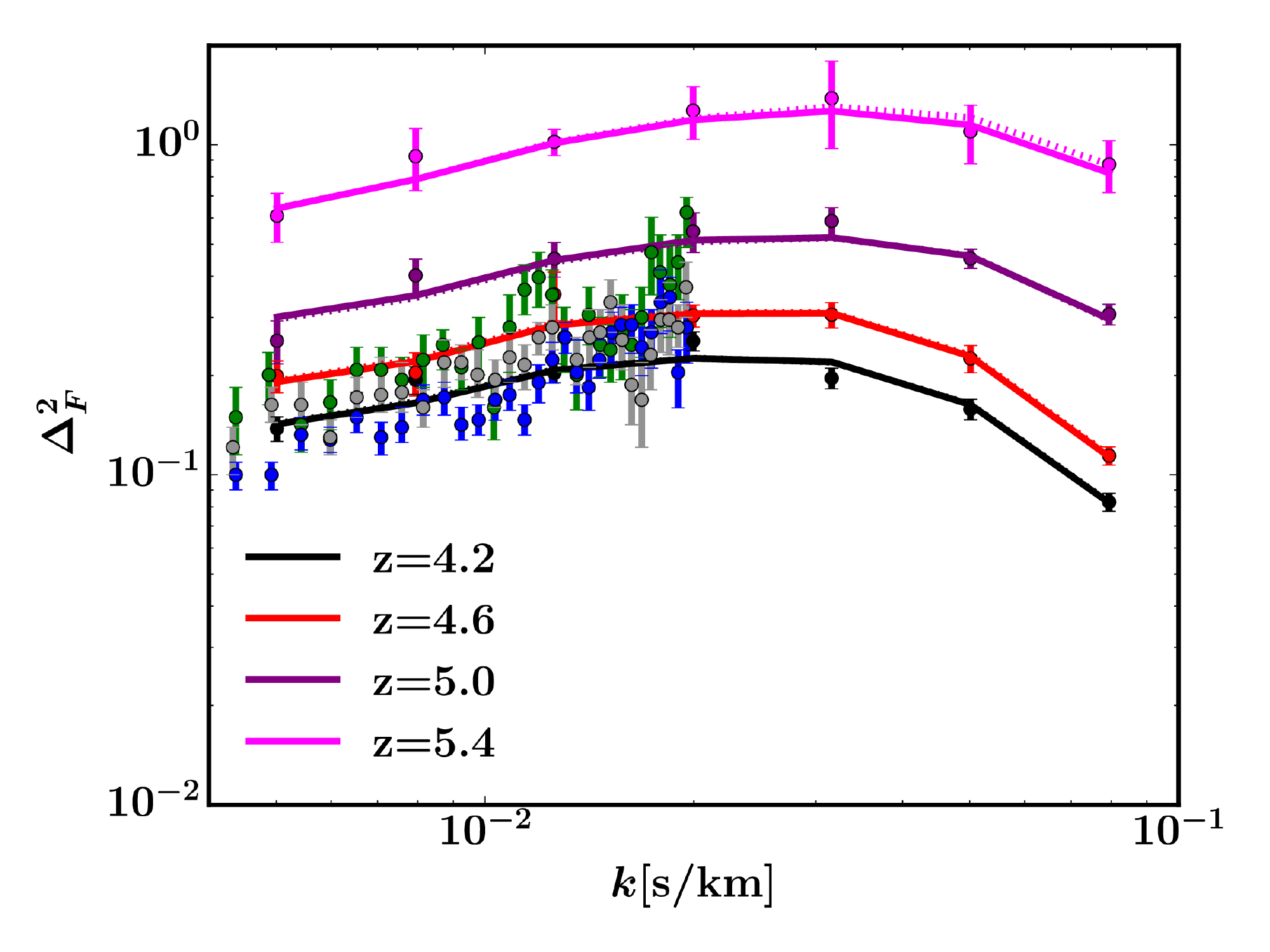}
  \caption{Measured flux power spectrum in dimensionless units,
    $\Delta^2_F(k)=P_F(k)\times k /\pi$, compared with the theoretical
    model with the best-fitting values of the astrophysical and
    cosmological parameters for WDM and CDM cosmologies. The solid
    refer the best-fitting values for WDM cosmology. The dotted lines
    refer to the best-fitting case for CDM cosmology. These
    best-fitting models largely overlap, except at the highest
    redshift and on the smallest scales. The blue, gray and green
    points are SDSS-III/BOSS DR9 data for $z = 4.0$, $z=4.2$ and $z =
    4.4$ from \citep{palanque2013}.}
  \label{fig:data}
\end{figure}

\begin{figure*}
  \centering \includegraphics[width=0.45\textwidth]{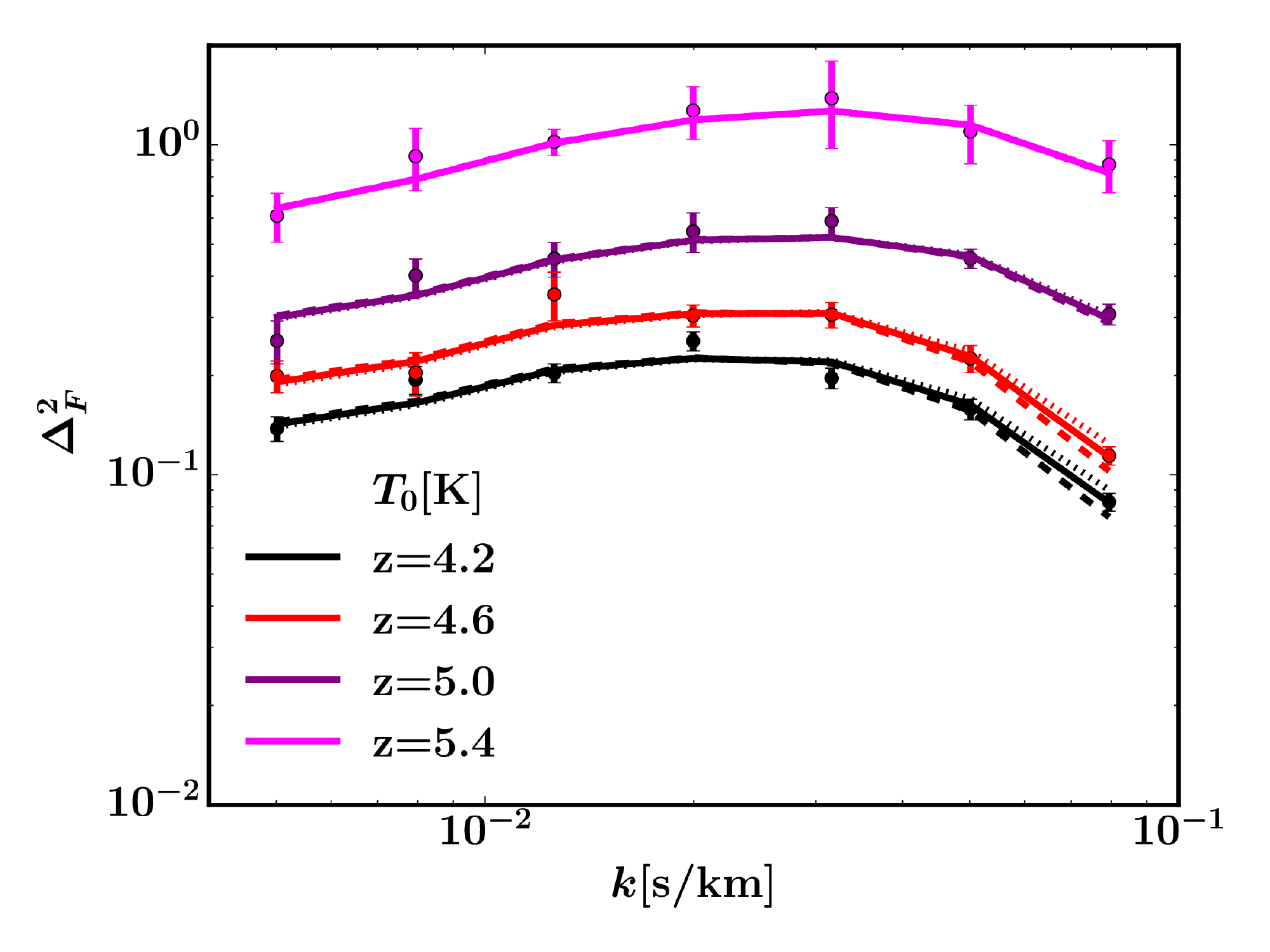}
  \centering \includegraphics[width=0.45\textwidth]{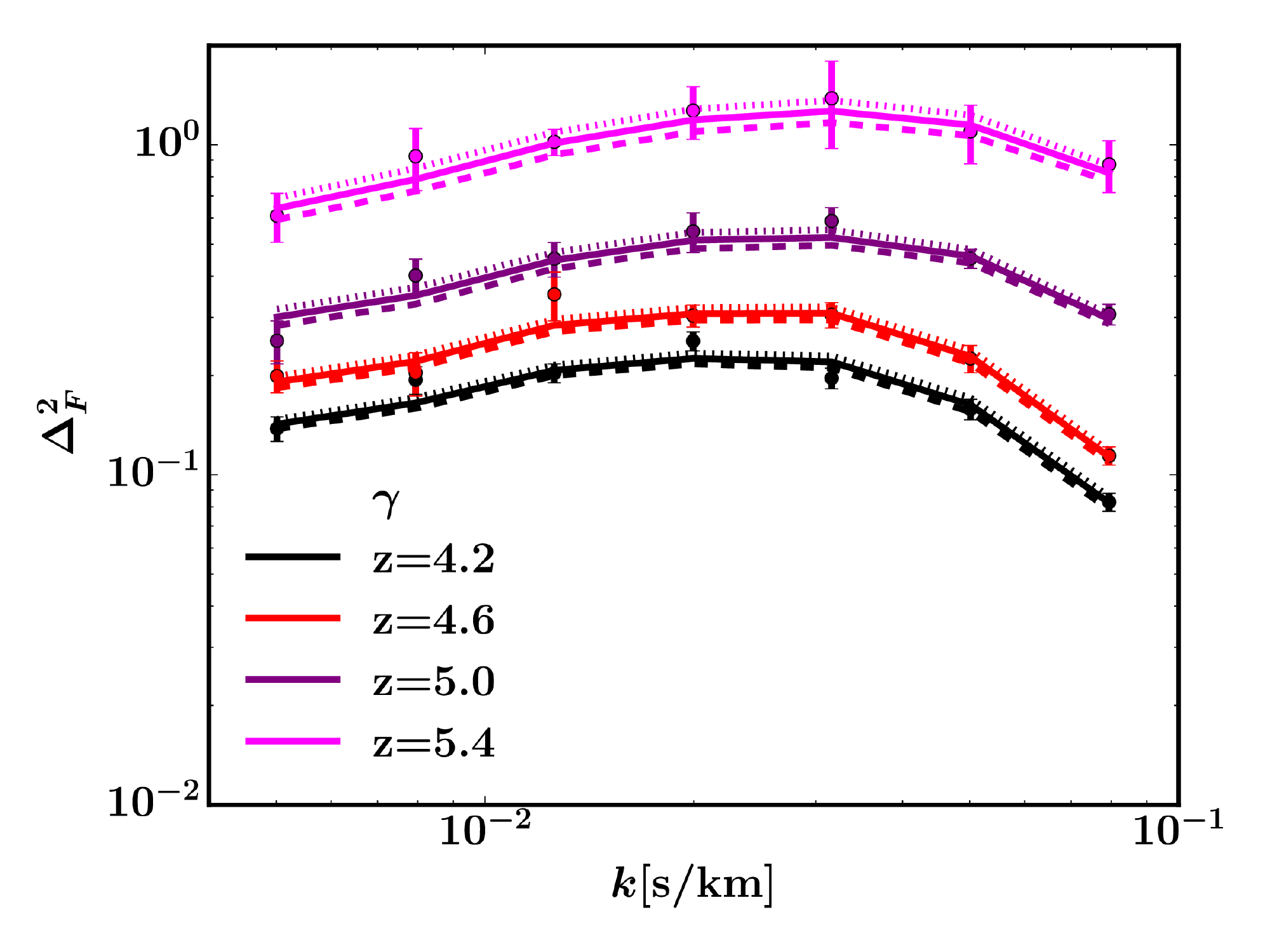}
  \centering \includegraphics[width=0.45\textwidth]{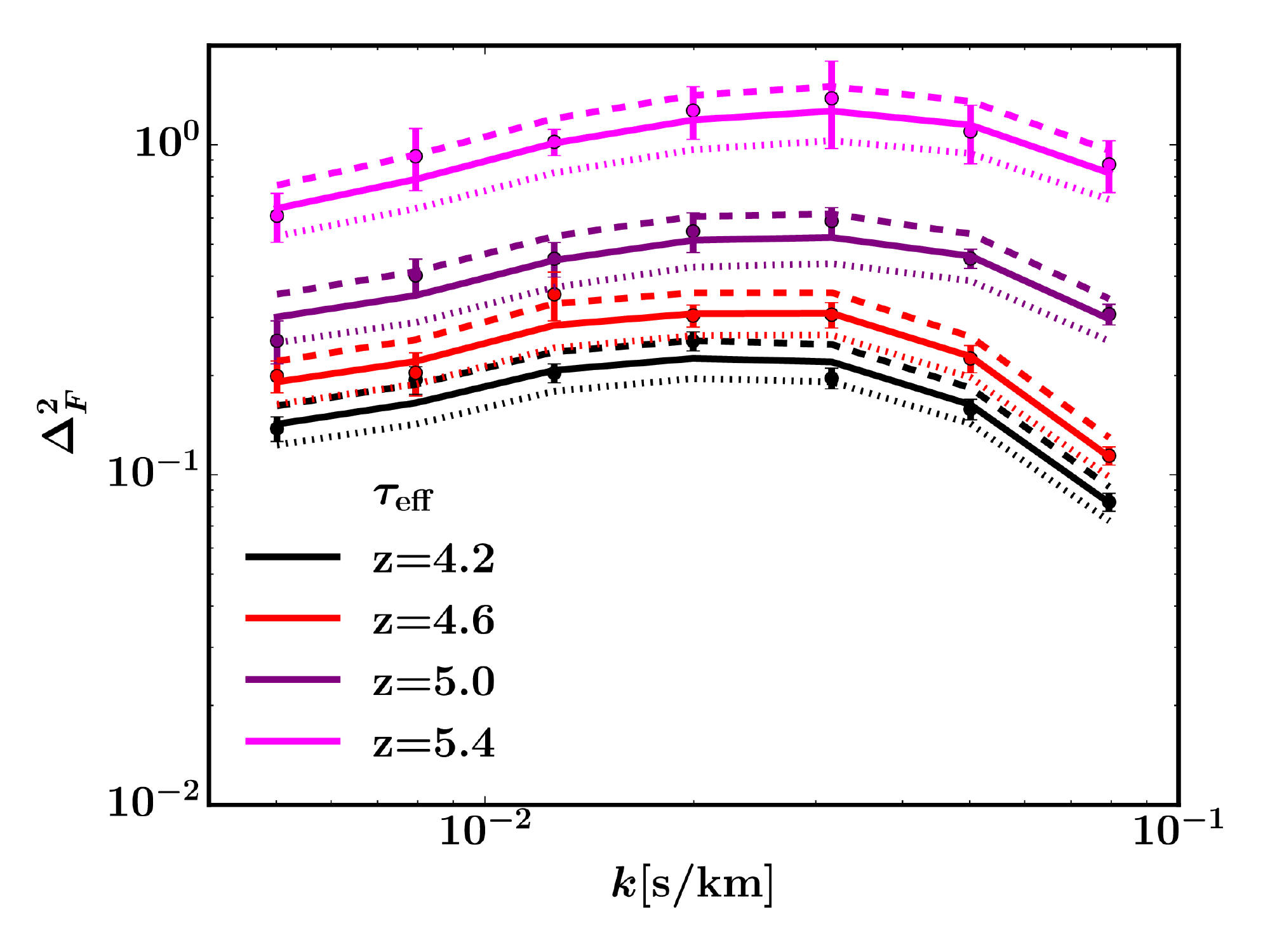}
  \centering \includegraphics[width=0.45\textwidth]{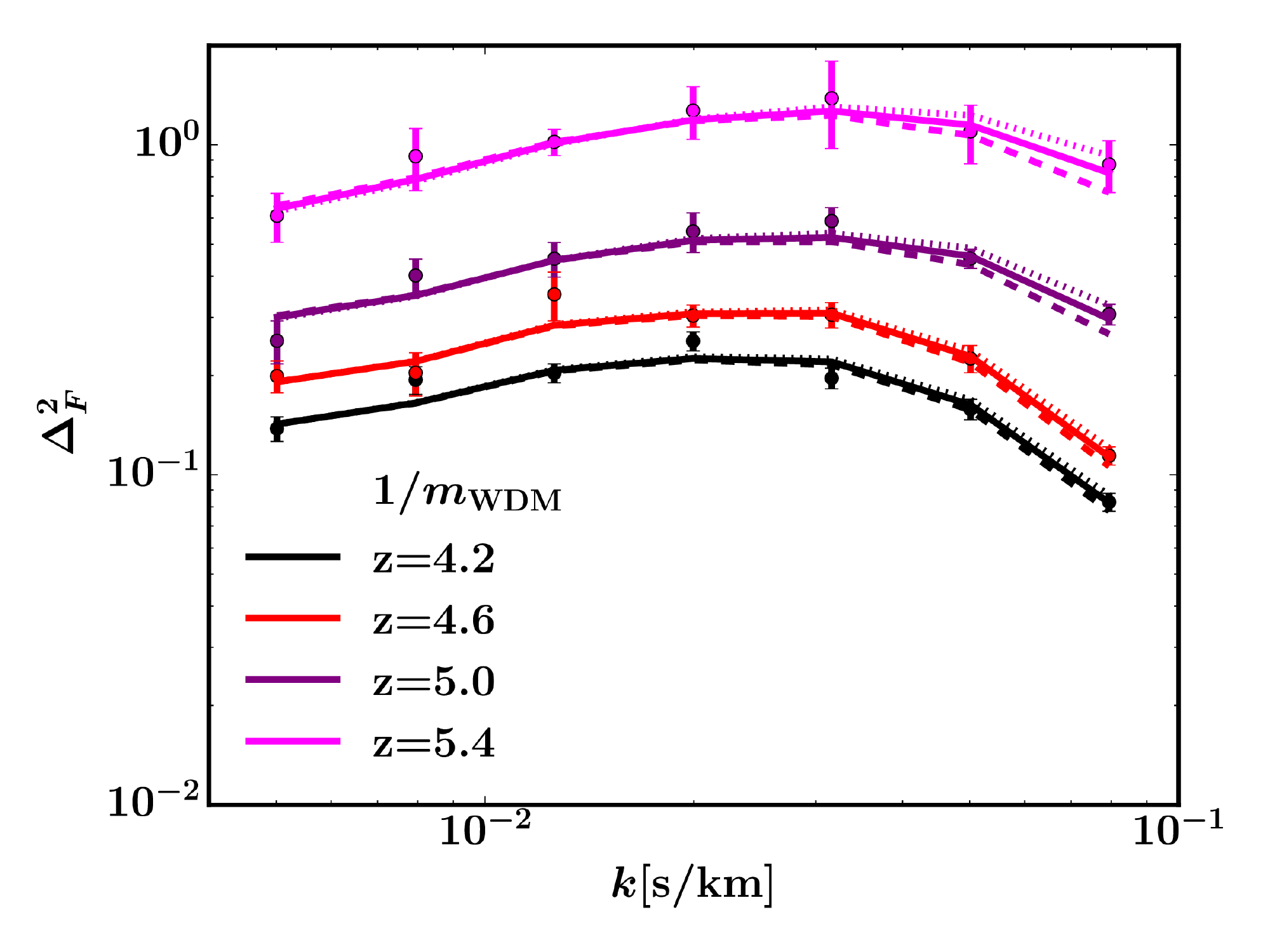}
  \caption{Effect of the IGM parameters and $m_\wdm$ on the flux power
    spectrum in dimensionless units, $\Delta^2_F(k)=P_F(k)\times k
    /\pi$. In the top-left (top-right, bottom-left, bottom-right)
    panel we show the effect of varying $T_0$ ($\gamma$, $\tau_{\rm
      eff}$, $1/m_\wdm$) by $\pm10\%$ with respect to the best-fitting
    values for WDM cosmology. The solid line corresponds to the
    best-fitting case for WDM cosmology, the dashed (dotted) line
    corresponds to the relevant parameter increased (decreased) by
    10\%.}
  \label{fig:changing}
\end{figure*}

\bigskip

\section*{Results}  In Table~\ref{tab:result} we give the result of the
parameter estimation. Fig.~\ref{fig:data} shows the theoretical flux
power spectrum for the mean values of the parameters, compared with
the MIKE and HIRES data used in this analysis.  In order to clarify
the effect of different thermal histories on our constraints, we show
the effect of changing the thermal parameters ($T_0$ and $\gamma$) and
ionization parameters ($\tau_{\rm eff}$) and the mass of the thermal
relic ($1/m_\wdm$)in Fig.~\ref{fig:changing}, analogous to Figs.~5 and
6 of \citep{Viel:2013apy}.

\begin{figure*}
  \centering \includegraphics[width=\textwidth]{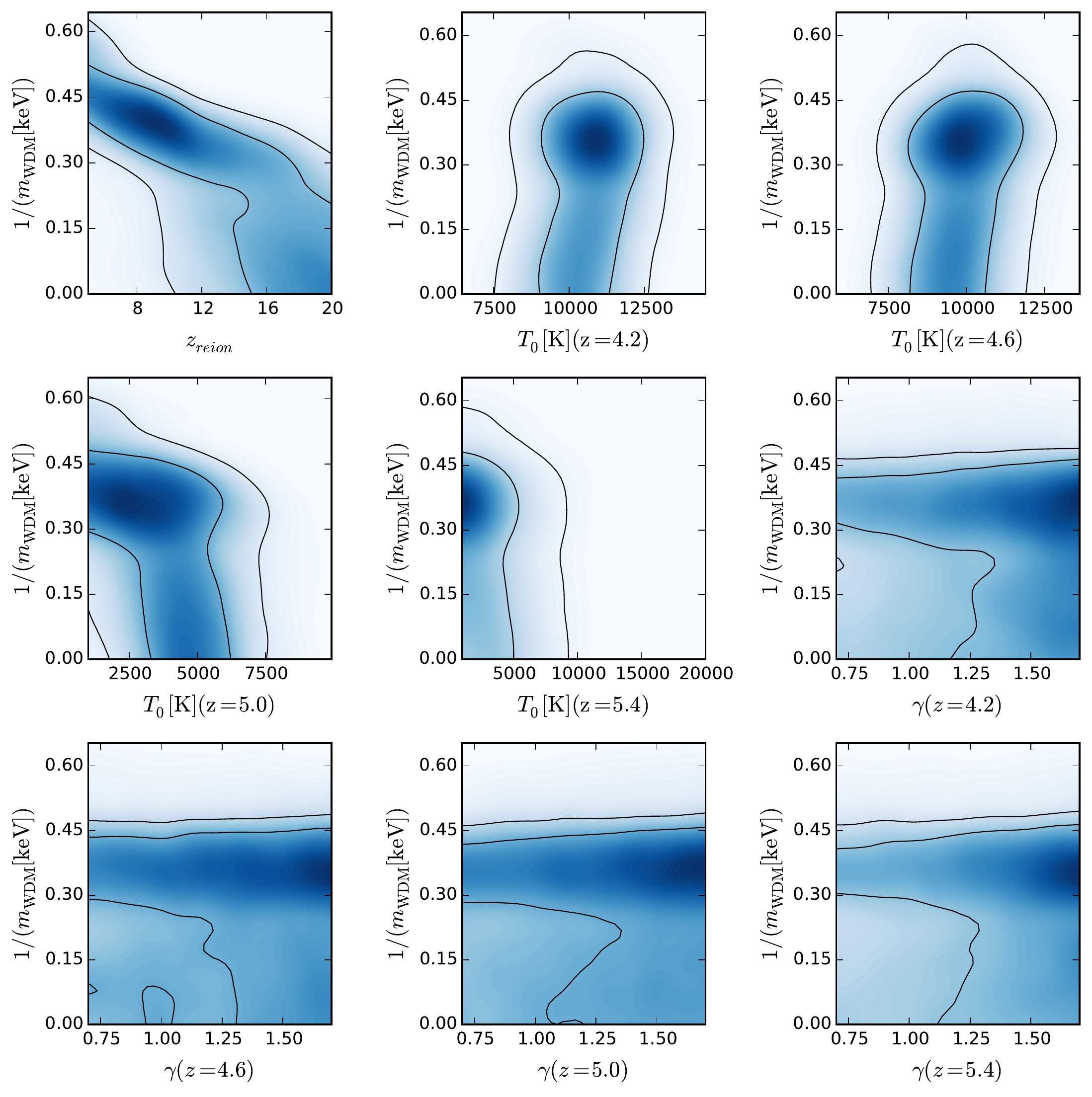}
  \caption{Confidence regions between $m_\wdm$, and $T_0$ and $\gamma$
    at all redshift, and $z_{\rm reion}$. We show $1/m_\wdm$ instead
    of $m_\wdm$ for visualization purposes. $m_\wdm$ is degenerate
    with $z_{\rm reion}$, that is the redshift at which the
    ultraviolet background has been switched on in the simulations,
    and $T_0$ at the redshift $z=5.0$. $m_\wdm$ is not degenerate with
    the $T_0$ for the other redshift intervals. There is no obvious
    degeneracy with $\gamma$.  Regarding $m_\wdm$ and $T_0$, at the
    redshifts $z=4.2,\; 4.6$ there is no degeneracy and $T_0 \sim
    10^4$~K is needed to explain the observed flux power spectrum,
    independently of $m_\wdm$.  At $z=5.0$ even very low temperatures
    $T_0 \lesssim 2500$~K are consistent with the data, and the cutoff
    in the flux power spectrum is explained by WDM rather than by the
    temperature. At $z=5.4$ the analysis prefers low values of $T_0
    \sim 5\times 10^3$~K, independently of $m_\wdm$.  }
  \label{fig:bayesian2D}
\end{figure*}

In Fig.~\ref{fig:bayesian2D} we show the 2D confidence regions between
$m_\wdm$, and $T_0\equiv T(\delta{=}0)$ (marginalizing over the other
parameters).  We see that at redshifts $z=4.2,\; 4.6$ there is no
degeneracy and an IGM temperature $T_0 \sim 10^4$~K is needed to
explain the observed flux power spectrum independently of $m_\wdm$.
If dark matter is ``too warm'' ($m_\wdm < 1.5$~keV) it produces too
sharp of a cut-off in the power spectrum and is inconsistent with the
data.

At the $z=5.0$ bin the situation is different. For the masses $m_\wdm \sim
2.2-3.3$~keV even very low temperatures $T_0 \lesssim 2500$~K are
consistent with the data. In this case the cutoff in the flux power
spectrum is explained by WDM rather than by the temperature. The
situation is analogous at $z=5.4$.  Table~\ref{tab:result} summarizes
the parameter estimation.

\begin{figure}
  \centering
  \includegraphics[width=\columnwidth]{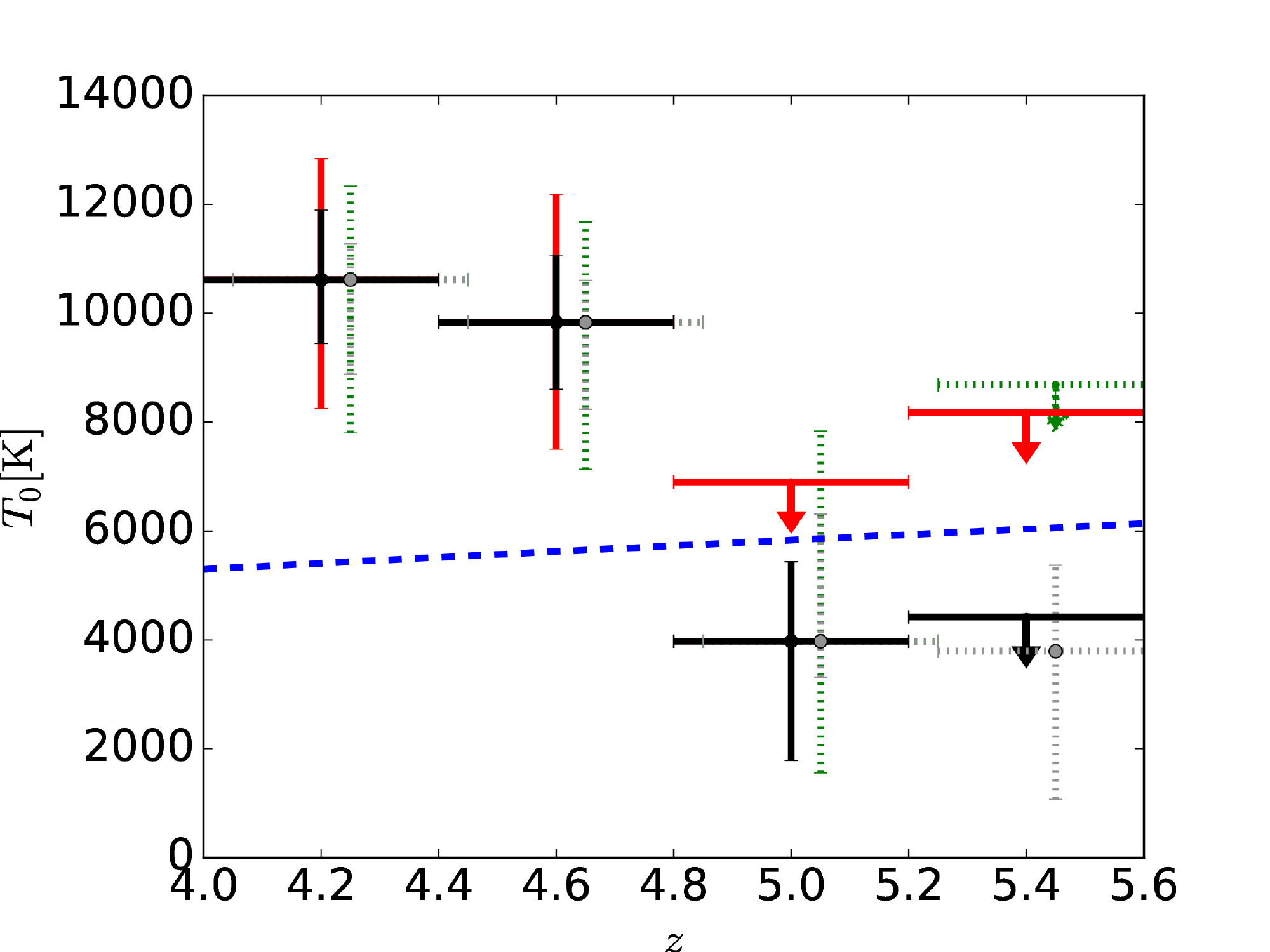}
  \caption{The evolution of the IGM mean temperature, $T_0$, in
    redshift. Black vertical bars are 1-$\sigma$ confidence limits;
    red vertical bars are 2-$\sigma$ confidence limits. Filled dots
    are the parameter mean; the arrows mark the upper limits. The
    horizontal bars span the redshift interval of Lyman~$\alpha$
    absorbers considered for each measurement of the flux power
    spectrum. The solid (dotted) lines refer to the constraints on
    temperature for WDM (CDM) cosmology (the constraints in CDM have
    been shifted by $z=0.05$ for improving the readability of the
    figure). At $z=5.0$ there is a 1-$\sigma$ level detection and only
    an upper limit at 2-$\sigma$ level in WDM cosmology, instead there
    are both 1 and 2-$\sigma$ detections for CDM cosmology. At
    $z=5.4$, there are only upper limits at 1 and 2-$\sigma$ levels
    for WDM cosmology and 1-$\sigma$ detection and 2-$\sigma$ upper
    limit for CDM cosmology. Hence, the constraints on the temperature
    are substantially equivalent in the two cosmologies. The blue
    dashed line is the asymptotic IGM mean temperature in the case of
    early hydrogen and first helium reionization from a stellar
    ionizing spectrum with slope $\alpha=2$, being the ionizing
    spectrum $J_{\nu} \propto \nu^{-\alpha}$.}
  \label{fig:t0z}
\end{figure}

Another important property of Fig.~\ref{fig:bayesian2D} is that even
assuming CDM cosmology, the temperature $T_0$ is a non-monotonic
function of redshift and should be colder than $\sim 8000$~K at
$z=5.0-5.4$, see Fig.~\ref{fig:t0z}.\footnote{The temperature values
  that we have estimated at high redshift could be inaccurate, because
  the lowest temperature in the simulation grid was $5400\,{\rm K}$.}

The resulting $\chi^2$ for the Bayesian analysis is $\sim 25$, for 30
degrees of freedom (49 data points - 19 free parameters). This is in
agreement with the fact that the covariance matrix is uncertain and
that has been multiplied by a factor that boosts the resulting error
bars by $30\%$, with respect to the error bars computed by
bootstrapping. This is done in the original analysis in order to
account for presumed sample variance effect that affect other
statistics like the transmitted flux PDF. The sample variance effect
may affect the transmitted flux power spectrum, even if a detailed
computation has not been performed.

\begin{figure}
  \centering
  \includegraphics[width=\columnwidth]{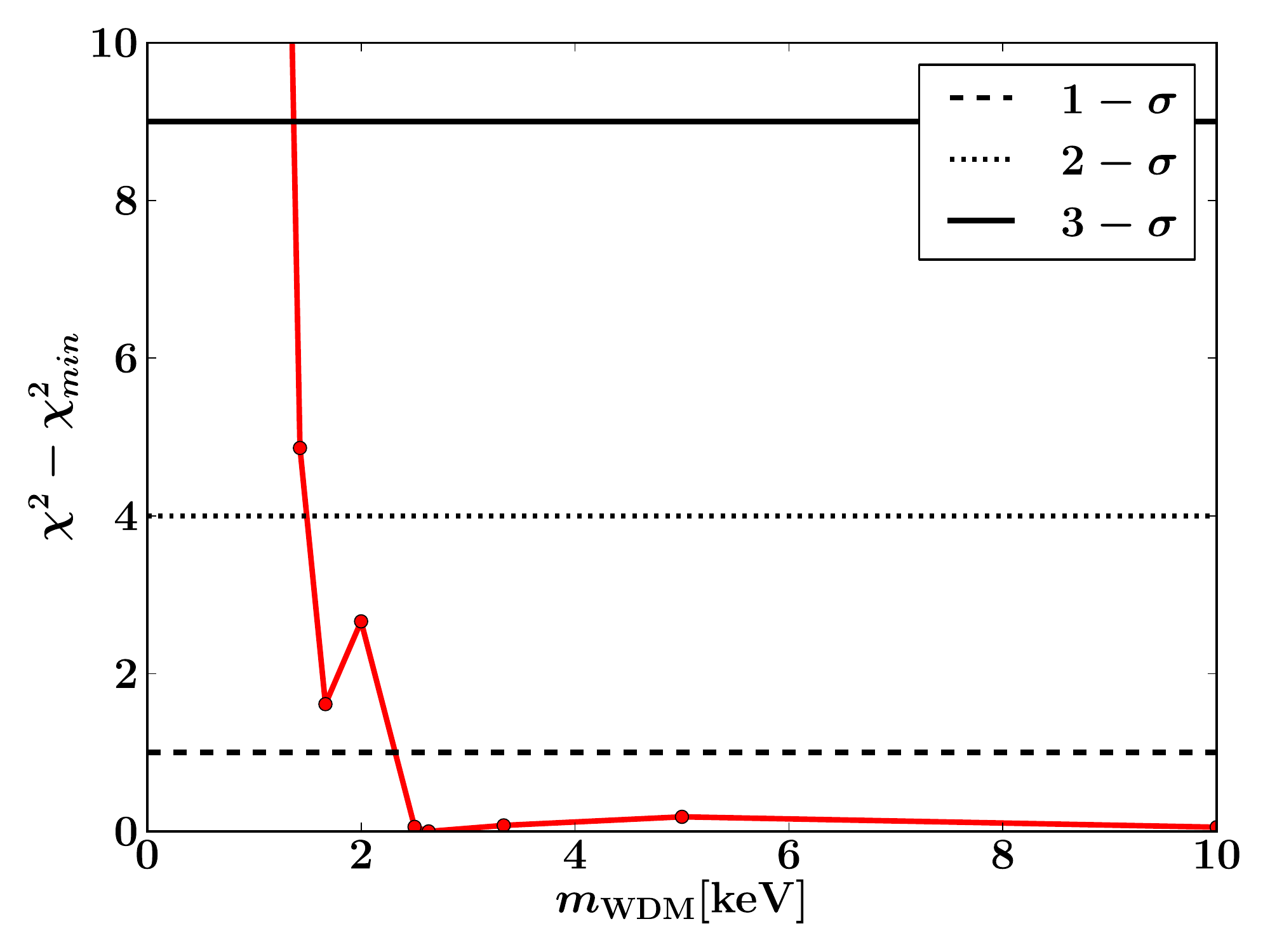}
  \caption{The results of the frequentist analysis: the
    $\chi^2-\chi_{min}^2$ versus the WDM mass, $m_{\rm WDM}$. There are two
    minima of the $\chi^2$ curve, CDM and $m_\wdm=2.7\,{\rm keV}$.}
  \label{fig:frequentist}
\end{figure}

For completeness we have also performed frequentist analysis for the
same $\chi^2$ considered in the Bayesian analysis. As shown in
Fig.~\ref{fig:frequentist} the two analyses are in broad agreement
with each other.

We would like to stress that our results depend crucially on allowing for a
non-monotonic redshift dependence of $T_0(z)$.  In~\cite{Viel:2013apy} it was
shown that assuming a power-law (\emph{monotonic}) redshift dependence for
$T_0(z)$ and $\gamma(z)$, one predicts higher temperatures of IGM for the same
data. In this case the CDM cosmology is preferred over WDM, leading to the
$2\sigma$ lower bound $m_{\wdm} \ge 3.3$~keV~\cite{Viel:2013apy}.  The
``binned analysis'' of~\cite{Viel:2013apy} gave results similar to those,
reported here. The authors of~\cite{Viel:2013apy} however rejected these
results, considering a temperature jump at $z=5-5.4$ to be ``unphysical'' and
arguing that the low $\chi^2$ is a sign of overfitting.

In our opinion the present analysis implies that more data is needed
to study such a scenario, as it currently does not allow to make any
definitive conclusion and in particular does not allow to rule it out.
Moreover, as mentioned above, the error bars in~\cite{Viel:2013apy}
were inflated by $30\%$ and therefore we consider the reduced $\chi^2
= 25/30 \approx 0.83$ to be consistent with $1$. We see that $2\sigma$
lower bound on the WDM mass relaxes down to $m_\wdm \ge 2.1$~keV
(consistently with the results of binned analysis
of~\cite{Viel:2013apy}). Moreover, the non-monotonic thermal history
makes the WDM with $m_\wdm = 2-3$~keV an equally good fit as CDM. The
best fit values of $T_0$ can be inaccurate as they lie below the
lowest simulation point in $T_0$ grid.  Therefore more simulations are
needed to settle this question. This is currently work in progress. In
the absence of such additional studies the proposed non-monotonic
thermal history cannot be ruled out based on the existing
Lyman-$\alpha$ data.

For the interpretation of these results it is crucial to overview what
is known about the thermal state of the IGM both theoretically and
observationally. We argue below that the measured thermal history is
in agreement with current models of galaxy formation and reionization.

\begin{table}
  \caption{Parameter estimation from Bayesian analysis. We show the
    1-$\sigma$ and 2-$\sigma$ confidence intervals. We only show the
    parameters that are constrained at 1 or 2-$\sigma$ level.}
  \small
  \label{tab:result}
  \begin{tabularx}{\columnwidth}{XX}
  \begin{tabular}{lccc}
    parameter & mean & 1-$\sigma$ & 2-$\sigma$\\
    \hline
    $H_0[{\rm km/s/Mpc}]$        & 63   & $<$  67       & --  \\
    $m_{\rm WDM}[{\rm keV}]$       & 3.9  & [143,   2.3]  & $>$ 2.1 \\
    $T_0(z=4.2)[10^3\,{\rm K}]$  & 10.6 & [9.4,   11.8] & [8.3,  12.9] \\
    $T_0(z=4.6)[10^3\,{\rm K}]$  &  9.8 & [8.6,   11.1] & [7.5,  12.2] \\
    $T_0(z=5.0)[10^3\,{\rm K}]$  &  4.0 & [2.0,   5.6]  & $<$ 6.9 \\
    $T_0(z=5.4)[10^3\,{\rm K}]$  &  3.8 & $<$ 4.5       & $<$ 8.2 \\
    $\tau_{\rm eff}(z=4.2)$        & 1.12 & [1.05,  1.19] & [1.00, 1.25] \\
    $\tau_{\rm eff}(z=4.6)$        & 1.30 & [1.21,  1.39] & [1.15, 1.47] \\
    $\tau_{\rm eff}(z=5.0)$        & 1.88 & [1.74,  2.00] & [1.64, 2.13] \\
    $\tau_{\rm eff}(z=5.4)$        & 2.91 & [2.69,  3.10] & [2.54, 3.31] \\
    $\gamma(z=4.2)$             &  1.3 & $>$ 1.1        & -- \\    
    $\gamma(z=5.4)$             &  1.3 & $>$ 1.1        & -- \\
  \end{tabular}
  \end{tabularx}
\end{table}

\bigskip 
\section*{State of the IGM at $z\sim 5$}  The IGM temperature
can be \emph{determined} from the broadening of the Lyman-$\alpha$
absorption lines in QSO spectra \citep{Theuns:2000va, McDonald:2000nn,
  Zaldarriaga:2000mz, Theuns:2000va, Viel:2005ha,Schaye:1999vr,
  Ricotti:1999hx, Lidz:2009ca, Becker:2010cu, bolton2012,
  Garzilli:2012gy, Rudie:2012mx, Garzilli:2015bha}. Alternatively, it
has been proposed to determine the IGM temperature by measuring the
level of the transmitted flux \citep{Bolton:2007xi, Viel:2009ak,
  Calura:2012qq, Garzilli:2012gy}, however there is no agreement
between the two methods yet, see \citep{rollinde2013}.

All the measurement of the IGM temperature in the literature assumed
CDM cosmology. Because of the existing degeneracy between the IGM
temperature and WDM, the assumption of the WDM cosmology could change
the deduced values of the IGM temperature. Nevertheless, in the
absence of such measurements, we compare our estimates for the IGM
temperature with the measurements based on the CDM assumption.

The
IGM temperature at $z<5$ is constrained relatively well to be at the
level $T_0\gtrsim(8-10)\times
10^3$~K~\cite{Schaye:1999vr,McDonald:2000nn,Lidz:2009ca,Becker:2010cu}.
At $z=6.0$ there is a single measurement, \citep{bolton2012}, that
restricts $T_0$ to the range $5000 < T_0 <10000$~K (68\% CL) (see
e.g.~\cite{Viel:2013apy} for discussion). The simplest interpretation
of these data (also adopted in \cite{Viel:2013apy}) is that the
temperature is growing monotonically with redshift.  Instead, given
the large error bars of the measurements, and taking into account
adiabating cooling one may expect a drop of temperature at $z\sim 5$
with a subsequent rise to $\sim 10^4$~K at $z\sim 4.6$ in agreement
with other measurements from
\cite{Schaye:1999vr,McDonald:2000nn,Lidz:2009ca,Becker:2010cu}. This
increase in IGM temperature can be explained with an early start of
HeII reionization predicted by some models of reionization by quasars,
\citep{mcquinn2009} (see recent discussion of such ``two-component''
reionization models in~\cite{Madau:2015cga}).

In such a scenario, the temperature at $5<z<6$ depends on how long the
first stage of reionization lasted and what the temperature of IGM was
at $z\gtrsim 6$.  As mentioned above, the
measurement~\cite{bolton2012} at $z\sim 6$ has large
uncertainties. Theoretically, $T_0(z{=}6)$ depends on how early the
first stage of hydrogen (and HeI) reionizations has ended, and what
sources drove it (c.f.\ \citep{haardt1996,hui2003}). It has been
speculated that hydrogen is reionized by the metal-free (Population
III) stars, whose spectral hardness predicts high values of the
temperature. However, the properties of Population~III stars are
purely speculative -- we do not know how long they lasted and whether
they were indeed the sources of reionization. For example,
reionization could be due to a more metal rich population of stars
with softer stellar spectra \citep{ciardi2005}, leading to a lower
values of IGM temperature at $z\sim 6$.  To settle this question, an
independent constraint on the ultraviolet background at high redshift
would be needed, however there are no such measurements to-date. The
lower limit of \citep{bolton2012} is $T_0(z{=}6)\approx 5\times
10^3$~K or even slightly below, fully consistent with the low values
at $z=5.0-5.4$ (Table~\ref{tab:result}) reached via adiabatic cooling.

We note that an indirect argument in favour of the IGM temperatures at
high redshifts being $\sim10^4\,{\rm K}$, is the ``missing satellite
problem'' -- high temperature would prevent gas from collapsing into
dark matter halos with a mass below $\sim 10^7 M_\odot$, thus
suppressing the formation of small galaxies (see
e.g.\ \citep{Benson:2001au,Benson:2001at,Maccio:10,sawala2014}),
explaining in particular the small number of satellites of the Milky
Way. However, in WDM cosmologies the matter power spectrum is
suppressed at the smallest scales, thus solving the missing satellite
problem even if the gas was sufficiently cooler.

\begin{figure}
  \centering \includegraphics[width=\columnwidth]{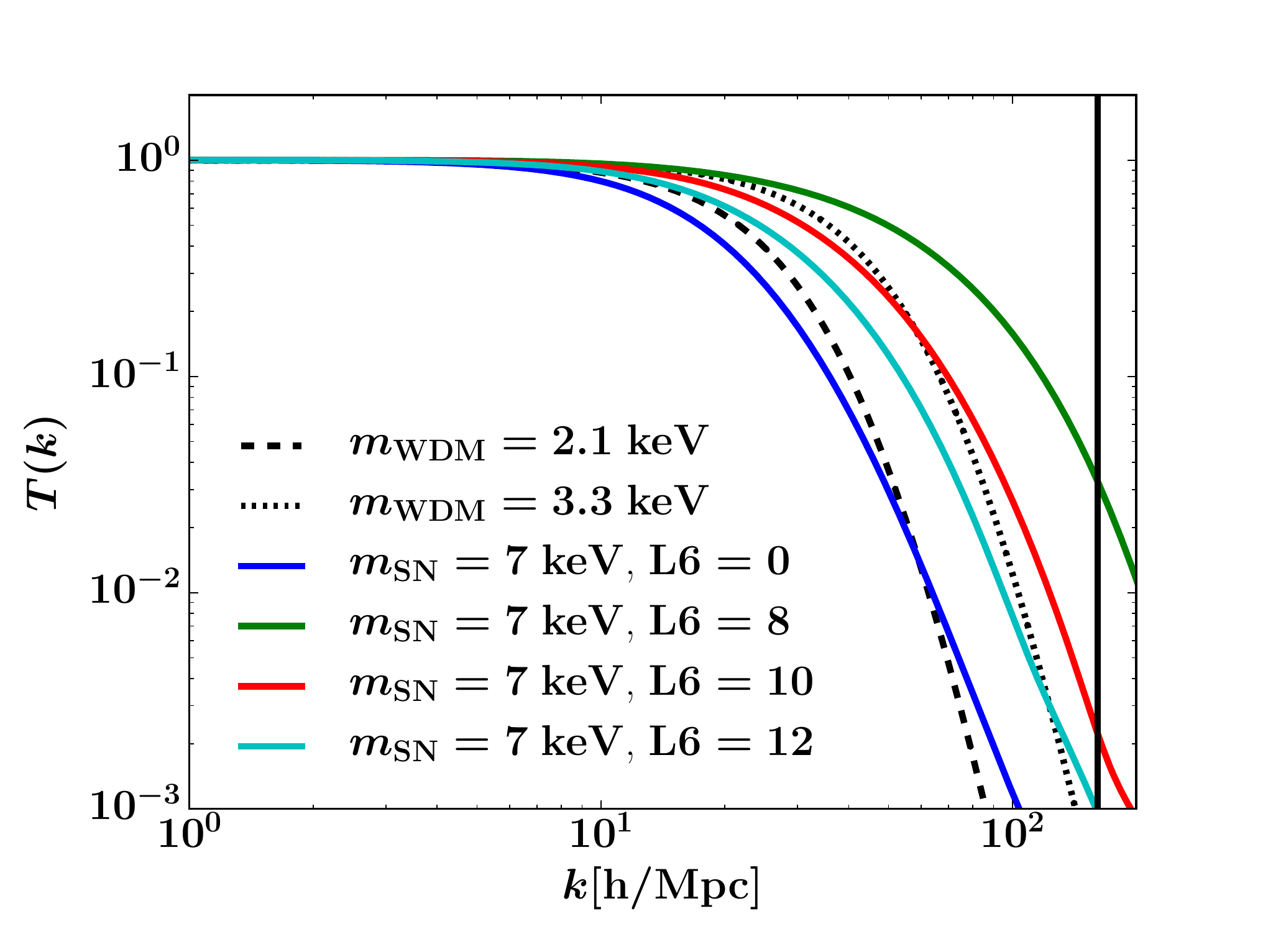}
  \caption{Comparison between the linear transfer functions, $T(k)$,
    of thermal relic (WDM) and sterile neutrinos (SN). The dashed
    (dotted) black line is the linear transfer function for $m_{\rm
      WDM}=2.1\, {\rm keV}$ ($m_{\rm WDM}=3.3\, {\rm keV}$) as
    computed in \citep{Viel:2005qj}. The colored (green, red, cyan)
    lines are realistic linear transfer functions for some of the
    sterile neutrino models with \mbox{$m^{\rm NRP}_{\rm SN}=7\, {\rm keV}$}.
    The linear transfer functions with ${L_6}=10$ and 12 (red and cyan
    lines) are partially warmer that the lower bound of
    \citep{Viel:2013apy} (the dotted black line), but still satisfy
    the constraints from this letter (the dashed black line) until the
    maximum $k$-mode used in the reference numerical simulations. The
    linear transfer function with $L_6=8$ (green line) is colder than
    the bound of \citep{Viel:2013apy}. The linear transfer function
    with ${L_6}=0$ (blue line) violates the constraint from this
    letter. The solid vertical line is the maximum $k$-mode used in
    the reference simulations.}
  \label{fig:SNspectra}
\end{figure}

Finally, we use our results to explore the constraints on sterile
neutrino dark matter~\cite{Dodelson:1993je}, resonantly produced in
the presence of lepton
asymmetry~\cite{Shi:1998km,Abazajian:01a,Laine:2008pg}. This is a
non-thermal warm dark matter, whose primordial phase-space density
distribution resembles a mixture of cold+warm dark matter
components~\cite{Boyarsky:2008mt,Boyarsky:2009ix}, demonstrating a
shallower cut-off. In Fig.~\ref{fig:SNspectra} we compare the linear
transfer function (the square root of the ratio of the modified linear
matter power spectrum to that of cold dark matter, $T(k) =
\sqrt{P_{\rm WDM}(k)/P_{\rm CDM}(k)}$) of thermal relic WDM with a
mass $m_\wdm = 2.1$~keV (lower bound from this work) and a $m_\wdm =
3.3$~keV~\cite{Viel:2013apy} with those of resonantly produced sterile
neutrinos with the mass 7~keV (motivated by the recent reports of an
unidentified spectral line at the energy $E\sim 3.5$~keV in the
stacked X-ray spectra of Andromeda galaxy, Perseus galaxy clusters,
stacked galaxy clusters and the Galactic Center of the Milky
Way~\cite{Bulbul:2014sua,Boyarsky:2014jta,Boyarsky:2014ska}). We show
that depending on the value of the lepton asymmetry, $L_6 \equiv
10^6(n_{\nu_e}- n_{\bar\nu_e})/s$ (see~\cite{Boyarsky:2009ix} for
details) the linear power can be colder than that of thermal relics
with $m_\wdm = 2.1$~keV (Fig.~\ref{fig:SNspectra}), thus being fully
admissible by the data.\footnote{Our computations of the phase-space
  distribution functions for sterile neutrinos are based
  on~\cite{Laine:2008pg} and the linear power spectrum is obtained
  with the modified CAMB code developed in~\cite{Boyarsky:2008mt}. We
  do not expect the most recent
  computations~\cite{Ghiglieri:2015jua,Venumadhav:2015pla} to affect
  our results.}  Notice that the non-resonant sterile neutrino dark
matter with a 7~keV mass would be excluded at more than $3\sigma$
level by previous constraints from the SDSS
\citep{Seljak:2006qw,Boyarsky:2008xj}.

\bigskip

\section*{Conclusion and future work}  We demonstrated that the cut-off in the flux
power spectrum, observed in the high resolution Lyman-$\alpha$ forest
data may either be due to free-streaming of dark matter particles or
be explained by the temperature of the intergalactic medium. Taking
into account measurements at redshifts $z\sim 6$ and at $z < 5$ we see
that if dark matter is \emph{warm}, this requires non-monotonic
dependence on the IGM temperature on $z$ with the local minimum at $z
\sim 5.0-5.4$.  Even cold dark matter slightly prefers a
non-monotononic $T_0(z)$.\footnote{As stated in \cite{Becker:2014oga},
  a model of fluctuating UVB, with spatially constant mean free path
  for the hydrogen-ionizing photons (similar to the one used in
  \cite{Viel:2013apy} and in this work) may not be adequate to explain
  the observed scatter in the optical depth at the redshifts $5.1\leq
  z\leq 5.7$. Proper modeling of patchy reionization may affect the
  conclusions about the IGM state. We leave this for future work.}
Improving our knowledge of the IGM temperature at $z \sim 5-6$ will
therefore either result in very strong Lyman-$\alpha$ bounds on DM
free-streaming, essentially excluding its influence on observable
small-scale structures, or (if temperature will be found to be well
below $5000$~K) would lead to the discovery of WDM.

A method that would allow to measure the IGM temperature at the
redshifts of interest was presented in \cite{Garzilli:2015bha}. It is
based on the following idea: for high resolution spectra it is not
necessary to study average deviation from the QSO continuum per
redshift bins (as it is done in lower resolution case) but it is
possible to identify individual absorption lines and to measure their
broadening.  The thermal Doppler effect broadens the natural
lorentzian line profile of the Lyman-$\alpha$ transition
proportionally to the square root of the temperature, and one would
like to use this information to determine the temperature of the IGM
\emph{directly}. However, there are other effects that contribute to
the line width --- the physical extent and the clustering of the
underlying filaments. The method of \cite{Garzilli:2015bha}
potentially allows to disentangle these effects.  In view of our
results it is important to attempt to apply this method to
observational data. This is a method that has been tested with
simulations at redshift $\sim 3$, and it still has to be seen if it
works at redshift 5.

\bigskip

\section*{Acknowledgments}  The authors are grateful to Matteo Viel,
for sharing with them the code used in \citep{Viel:2013apy}, and
making this analysis possible.  The authors thank James Bolton, Joop
Schaye and Tom Theuns for useful discussions on the IGM temperature at
high redshift.  AG thanks Mark Lovell for sharing his knowledge about
generating SN power spectra. She also thanks Bin Hu, Samuel Leach,
Matteo Martinelli and Jesus Torrado for useful discussions on MCMC
method. The authors thank the anonymous referee for his helpful
comments, that improved the manuscript by large extent. The research
was supported in part by the European Research Council under the
European Union's Seventh Framework Programme (FP7/2007-2013)/ERC grant
agreements 278594-GasAroundGalaxies.

\section*{References}

\bibliography{apstemplate}

\end{document}